\documentstyle[11pt,paspconf,epsf]{article}
\newcommand{\mc}{\multicolumn}

\newcommand{\lta}{\stackrel{<}{_{\sim}}}

\begin{document}

\title{No evidence for a `redshift cut-off' for the most powerful
classical double radio sources}
\author{Matt J.\,Jarvis$^{1}$, Steve Rawlings$^{1}$, Chris J.\
Willott$^{1,2}$, Katherine~M.\,Blundell$^{1}$, Steve Eales$^{3}$,
Mark Lacy$^{1}$}
\affil{$^{1}$Astrophysics, Department of Physics, Keble Road, Oxford, OX1 3RH. \\
$^{2}$Instituto de Astrof\'\i sica de Canarias, 
C/ Via Lactea s/n, 38200 La Laguna, Tenerife, Spain \\
$^{3}$Department of Physics and Astronomy, University of Wales 
College of Cardiff, P.O. Box 913, Cardiff, CF2 3YB\\}

\begin{abstract}
We use three samples (3CRR, 6CE and
6C*) to investigate the radio luminosity function (RLF) for the `most
powerful' low-frequency selected radio sources. 
We find that the data are well fitted by a model with a constant
co-moving space density at high redshift as well as by one with a
declining co-moving space density above some particular redshift.
This behaviour is very similar to that inferred for
steep-spectrum radio quasars by Willott et al\,(1998) in line with
the expectations of Unified Schemes. We conclude that there is as yet
no evidence for a `redshift cut-off' in the co-moving space densities of
powerful classical double radio sources, and rule out a cut-off at $z\lta2.5$.
\end{abstract}
\section{Introduction}
It is well-known that the (co-moving) space densities of the rarest,
most powerful quasars and radio galaxies were much higher at epochs
corresponding to $z \sim 2$ than they are now (Longair 1966). The
behaviour of the space density beyond these redshifts is the subject
of this paper. Dunlop \& Peacock (1990) found evidence for a `redshift
 cut-off' (a decline in the co-moving space density) in the distribution of flat-spectrum radio sources over the
redshift range $2-4$.  Through failing to find any flat-spectrum radio
quasars at $z >5$ in a large ($\approx\,$40 per cent of the sky) survey,
Shaver et al (1996, hereafter SH96) argued for an order-of-magnitude
drop in space density between $z \sim 2.5$ and $z \sim 6$, for this
class of object.  As emphasised by SH96, the crucial advantage of any
radio-selected survey is that with sufficient optical follow-up, it
can be made free of optical selection effects, such as increasing dust
obscuration at high redshift.  It is chiefly for this reason that the
work of SH96 provides the most convincing evidence to date for the
existence of an intrinsic decline in the co-moving space density of
any galaxy class at very high redshift.
\begin{figure}[!h]
\plotfiddle{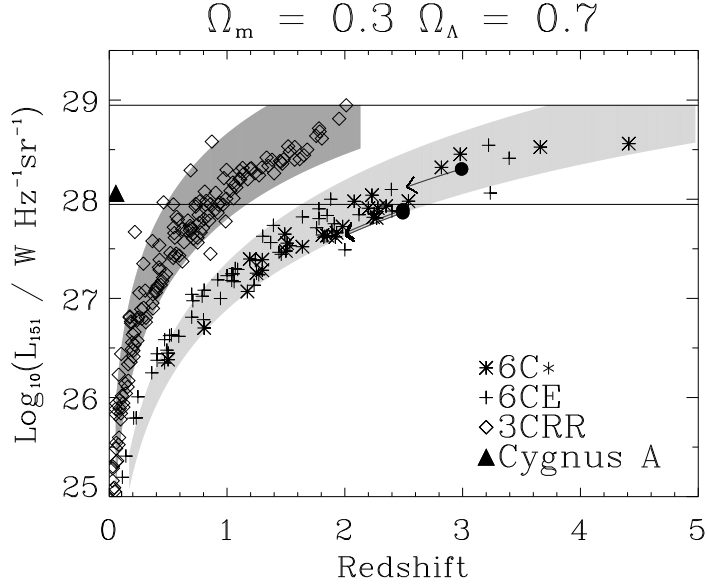}{3.1cm}{0}{80}{80}{-250}{-400}
\vspace{2.8cm}
{\caption{\label{fig:rad_1} $L_{151}-z$ plane for the three samples
used in our analysis. 3CRR (diamonds; Laing, Riley \&
Longair 1983), 6CE (crosses; Eales et al 1997), 6C* (stars; Blundell
et al 1998; Jarvis et al in prep.) and Cygnus-A (filled
triangle). The filled circles represent the three 6C* sources with
upper redshift limits and the associated arrows show the paths of the
possible redshifts and luminosities for these sources. The area
between the horizontal lines is the region which contains the `most
powerful' sources according to our definition.  The dark grey region
shows the approximate $L_{151}-z$ plane coverage of the 3CRR sample
($10.9\:{\rm Jy} \leq S_{178} \leq 80.0\:{\rm Jy}$, spectral index,
$\alpha = 0.5$) and the light grey region the 6C* sample ($0.96\:{\rm
Jy} \leq S_{151}
\leq 2.0\:{\rm Jy}$,
with $\alpha = 0.98$ and $\alpha = 1.5$ for the lower and upper fluxes,
 respectively). Note the area between the two shaded regions contains
no sources, this is the area which corresponds to the absence of a
flux-limited sample between the 6CE ($S_{151} \leq 3.93\:$Jy) and 3CRR
($S_{178} \geq 10.9\:$Jy) samples.}}
\end{figure}
\section{Modelling the RLF}
We adopt a parameterisation of the RLF which is separable in 151-MHz
luminosity $L_{151}$ and redshift $z$ with a single power-law in
$L_{151}$ of the form $(L/L_{\circ})^{-\beta}$. We consider two
cosmologies $\Omega_{\rm M}=1$, $\Omega_{\Lambda}=0$ (cosmology I) and
$\Omega_{\rm M}=0.3$, $\Omega_{\Lambda}=0.7$ (cosmology II).  Model A
parameterises the redshift distribution as a single power-law of the
form $(1+z)^{n}$. For model B the redshift distribution is
parameterised as a Gaussian, giving an overall expression for the
co-moving space density of $\rho = \rho_{\circ} (L/L_{\circ})^{-\beta}
\exp-\{(z-z_{\circ})/\sqrt{2}\,z_{1}\}^{2}$ where $\rho_{\circ}$ and
$\beta$ are the normalising term and power-law exponent respectively,
$z_{\circ}$ is the Gaussian peak redshift and $z_{1}$ is the
characteristic width of the Gaussian. Model C is described by the same
model up to $z_{\circ}$ beyond which it becomes constant.
\section{Results and Discussion}
For sources in the top-decade in luminosity of the $L_{151}-z$ plane
(Fig.\,1) our parametric fitting and likelihood analysis of model
radio luminosity functions (Table 1) show that the data are
inconsistent with a $(1 + z)^n$ power-law in redshift (Model A), but
are well fitted by both models B and C. These models are shown in
Fig.$\,2$ in the form of a log$\,N$ / log$\,S$ plot.  We conclude that
although the relative likelihood for model B is $2.5$ times larger
than for model C, this is not statistically significant enough to
distinguish between the two models with any confidence. This
uncertainty is further compounded by the effects of assuming a mean
spectral index in the model fitting. This result is in very close
agreement with the RLFs for radio loud quasars modelled by Willott et
al (1998) and various studies of AGN at optical (Irwin et al 1991) and
X-ray (Hasinger et al 1998) wavelengths.

This is in apparent contradiction to the findings of SH96 for
flat-spectrum quasars. If the relationship between the flat- and
steep-spectrum populations is as described by unification models of AGN
then we might expect to see similar evolution in the two populations. Thus to
determine the co-moving space density of radio sources at high-redshift,
an understanding of the spectral index trends, $K-$corrections and
associated selection effects must first be achieved. 

Fig.$\,2$ also illustrates the contribution of powerful sources at high
redshift to the total source count in a low-frequency
survey. We see that even for the no cut-off model (Model C) the
fractional contribution is very small. This may render the
location of the redshift cut-off virtually impossible to determine until the
selection effects associated with radio surveys are fully understood.
\begin{table*}
\small
\begin{center}
\begin{tabular}{cccccccll}
\hline\hline
\mc{1}{c}{Model} & \mc{1}{c}{Cosmology} & \mc{1}{c}{$\log_{10}(\rho_{\circ})$} & \mc{1}{c}{$\beta$} & \mc{1}{c}{$n$} & \mc{1}{c}{$z_{\circ}$} & \mc{1}{c}{$z_{1}$} & \mc{1}{c}{$P_{KS}$} & \mc{1}{c}{$\mathcal{L}_{R}$} \\
\hline\hline
A & I & $-9.04$ & 1.61 & 1.19 & ----- & ----- & 0.10 & $10^{-5}$ \\
B & I & $-7.94$ & 1.98 & ----- & 2.59 & 0.94 & 0.33 & 1 \\
C & I & $-8.18$ & 1.95 & ----- & 1.69 & 0.54 & 0.41 & 0.4 \\
\hline
A & II & $-9.45$ & 1.63 & 0.85 & ----- & ----- & 0.12 & $10^{-5}$ \\
B & II & $-8.51$ & 2.00 & ----- & 2.60 & 0.96 & 0.36 & 1 \\
C & II & $-8.78$ & 1.93 & ----- & 1.67 & 0.53 & 0.41 & 0.3 \\
\hline\hline
\end{tabular}
\end{center}
{\caption[junk]{\label{tab:fignotes} Best-fit parameters
for the model RLFs, described in the text. $P_{KS}$ is the 2-D
Kolmogorov-Smirnov probability (a value above $0.2$ signifies a
reasonable fit to the data) and
$\mathcal{L}_{R}$ is the likelihood relative to model B. \\
Errors for model B (cosmology I):
$\Delta\log_{10}(\rho_{\circ}) = 0.17$, $\Delta\beta = 0.2$,
$\Delta z_{\circ} = 0.10$, $\Delta z_{1} = 0.17$, for 68\% confidence
regions.}}
\end{table*}
\begin{figure}[!hb]
\plotfiddle{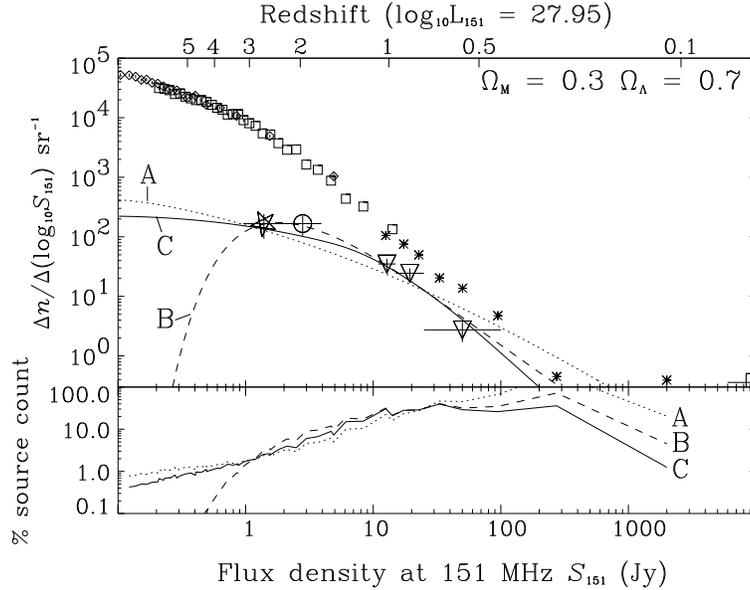}{3.1cm}{0}{53}{53}{-170}{-195}
{\caption[junk]{\label{fig:rad_2} For cosmology II we show the areal
density of the most powerful radio sources as a function of 151-MHz
flux density $S_{151}\,/\;$Jy in comparison with the total source
count. The 6CE source count (squares) is reproduced from Hales,
Baldwin \& Warner (1988) and the 7C source count (diamonds) from
McGilchrist et al (1990). The 3C source (stars) count was inferred
from the revised 3CR sample of Laing, Riley \& Longair (1983). The
large open star represents the 6C* data (note that this is a
lower-limit due to spectral index and angular size selection in this
filtered sample); the open circle the 6CE sample; the inverted triangles the 3
bins representing the 3CRR sample. The solid horizontal lines show the
$S_{151}$ range of each sample. The lower panel shows the percentage
of the total source count contributed by each model. Models A, B and C
are represented by the dotted, dashed and solid lines respectively. }}
\null
\end{figure}


\begin{references}
\reference Blundell K.M., Rawlings S., Eales S.A., Taylor G.B.,
Bradley A.D., 1998, MNRAS, 295, 265 
\reference Dunlop, J.S. \& Peacock, J.A., 1990, MNRAS, 247, 19 
\reference Eales S.A., Rawlings, S., Law-Green, D., Cotter, G., Lacy,
M., 1997, MNRAS, 291, 593 
\reference Hales, S.E.G., Baldwin, J.E. \& Warner, P.J., 1988, MNRAS,
234, 919
\reference Hasinger, G., 1998, Astron. Nachr., 319, 37
\reference Irwin, M., McMahon, R. \& Hazard, C., 1991, in The Space
Distribution of Quasars, ed. Crampton, ASPCS 21, 117
\reference Laing, R.A., Riley, J.M. \& Longair, M.S., 1983, MNRAS,
204,151 
\reference Longair, M.S., 1966, MNRAS, 133, 421 
\reference McGilchrist M.M., Baldwin, J.E., Riley, J.M., Titterington, D.J.,
Waldram, E.M., Warner, P.J., 1990, MNRAS, 246, 110
\reference Shaver, P.A., Wall, J.V., Kellermann, K.I., Jackson, C.A.,
Hawkins, M.R.S., 1996, Nature, 384, 439
\reference Willott, C.J., Rawlings, S., Blundell, K.M., Lacy M.,
1998, MNRAS, 300, 625
\end{references}
\end{document}